\documentclass[manuscript]{acmart}
\AtBeginDocument{%
  }

\usepackage{listings}
\usepackage[most]{tcolorbox}
\usepackage{tabularx}
\setcopyright{acmlicensed}
\copyrightyear{2018}
\acmYear{2018}
\acmDOI{XXXXXXX.XXXXXXX}
\acmConference[Conference acronym 'XX]{Make sure to enter the correct
  conference title from your rights confirmation email}{June 03--05,
  2018}{Woodstock, NY}
\acmISBN{978-1-4503-XXXX-X/2018/06}

\begin{document}

\title{Vulnerabilities in Personalization: Assessing Health Privacy Risks in ChatGPT Logs and Memory}

\author{S M Mehedi Zaman}
\email{sm.mehedi.zaman@rutgers.edu}
\author{Md Mozammel Hoque}
\email{mh1764@scarletmail.rutgers.edu}
\affiliation{%
  \institution{Rutgers University}
  \city{New Brunswick}
  \state{New Jersey}
  \country{USA}
}








\renewcommand{\shortauthors}{Trovato et al.}

\begin{abstract}
As conversational LLMs become deeply embedded in daily life, users frequently disclose sensitive personal health information during routine interactions. We present a large-scale computational audit analyzing 179,057 conversations across India, Nigeria, Brazil, and Pakistan ($N = 1,057$) to evaluate personal health disclosures and background memory synthesis in ChatGPT. We find that 21.31\% of audited conversations contain personal health data, with 3.62\% posing high-to-extreme privacy risks involving stigmatized conditions, direct identifiers, and precise locations. When evaluating the memory entries of ChatGPT, we uncover a stark disconnect between corporate framing and system behavior: over 95\% of profile entries are implicitly extracted without explicit user prompts or consent. Furthermore, background memory synthesis selectively condenses temporary, symptom-level disclosures into permanent diagnostic traits, stripping contextual integrity and amplifying re-identification risks. We conclude with sociotechnical design guidelines to restore user agency and consent-driven boundaries in stateful AI systems.
\end{abstract}

\ccsdesc[500]{Security and privacy~Social aspects of security and privacy}


\keywords{Health privacy, Algorithmic memory, Global south, Parasocial Relationship, ChatGPT.}

\received{20 February 2007}
\received[revised]{12 March 2009}
\received[accepted]{5 June 2009}

\maketitle

\section{Introduction}

\begin{figure}[htbp]
    \centering
    \includegraphics[width=\columnwidth]{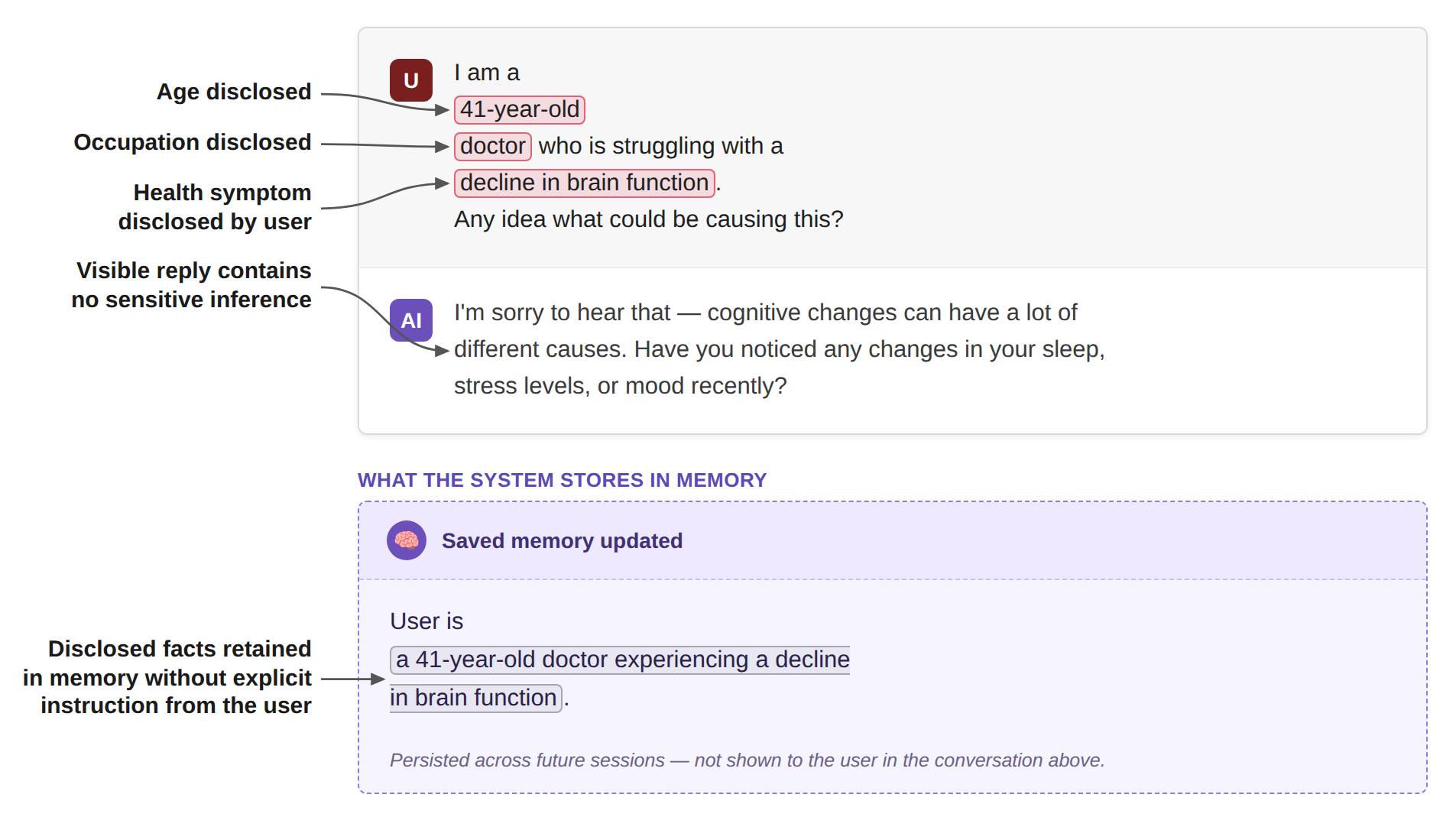}
    \caption{ A fictional example of a persistent-memory privacy risk in an LLM-based health conversation, inspired by our real-world dataset: a user disclosed their age, occupation, and a cognitive symptom. Three issues are demonstrated in the example: 1. the user discloses demographic information and a health symptom in their message; 2. the assistant's visible reply contains no explicit sensitive inference, giving the impression that nothing sensitive was retained; 3. despite this, the system's saved memory silently retains the disclosed facts without explicit instruction from the user to do so, persisting them across future conversations without the user's awareness or consent.}
    \label{fig:memory_retention}
\end{figure}

\begin{quote}
\textit{``Dreaming leverages a background process that allows ChatGPT to learn from many conversations and synthesize ChatGPT’s memory state in order to always provide the freshest, most relevant context... knowing you, helping you, and doing more for you.''} 
\hfill --- OpenAI (\textit{Dreaming: Better memory for a more helpful ChatGPT}, 2026) \cite{openai_dreaming_2026}
\end{quote}
When OpenAI announced its upgraded memory architecture built on background ``dreaming'' \cite{openai_dreaming_2026}, the value proposition was framed around seamless personalization: eliminating repetitive context-setting so users never have to introduce themselves from scratch. Rather than relying on manual instructions to save facts, the system operates continuously behind the scenes to synthesize long-term profile states from natural dialogue. Yet, while OpenAI illustrates this capability using benign preferences like travel plans or camera gear, applying automated background extraction to private personal disclosures introduces a profound, often unexamined privacy trade-off. When users engage with conversational AI chatbots, they frequently reveal sensitive personal contexts (e.g. figure~\ref{fig:memory_retention}) under the assumption that their input remains confined to an isolated chat thread \cite{zhang2024s}. In reality, background memory architectures quietly extract, condense, and persist these private disclosures across sessions without real-time user confirmation or explicit friction.
This tension between background memory extraction and thread-bound user expectations becomes especially dangerous in sensitive domains like personal health, particularly across the Global South \cite{gopichandran2019ethical}. In regions like India, Pakistan, Brazil, and Nigeria, millions of users navigate severe structural healthcare barriers—including physician shortages, overcrowded public clinics, and high out-of-pocket medical costs \cite{weissglass2022contextual}. In these resource-constrained settings, low to zero-cost consumer LLMs serve as informal ``digital doctors'' for immediate self-triage \cite{maity2025large}. Driven by empathetic interface cues and the ``trust-your-doctor'' heuristic \cite{wegwarth2013trust}, users share rich personal details, including diagnostic symptoms, financial distress, and local geography \cite{khan2025cross}. However, a critical research problem lies in the disconnect between user mental models and system mechanics: while users view these sensitive health exchanges as temporary consultations, ChatGPT's memory engine implicitly converts them into static, persistent profile traits. Despite corporate documentation framing memory as a user-controlled feature in the earlier ``legacy memory'' \cite{openai_memory_faq_2026}, it remains unknown how much health data is implicitly extracted, how severe the privacy risks are, and how condensed memory entries alter privacy risk compared to raw chat logs.

To address this gap, we present a large-scale computational audit evaluating the privacy mechanics of persistent LLM memory systems (specifically ChatGPT) using around 180,000 real-world conversation logs from four Global South countries (India, Pakistan, Brazil, and Nigeria), across more than a thousand users. We investigate three central research questions: \textit{RQ1: What sensitive health disclosures do users from the Global South share with ChatGPT during conversational interactions, and how severe are the associated privacy risks?} \textit{RQ2: How does ChatGPT's persistent memory extract and aggregate health disclosures, and how do the privacy risks of condensed memory profiles compare to raw conversational logs?} and \textit{RQ3: Does ChatGPT's persistent memory operate under explicit user direction as stated in official product documentation, or does it implicitly extract and save private health information without user agency?} Our audit reveals that health triage queries account for a substantial fraction of user interactions, frequently involving direct patient identifiers, geographic locations, and stigmatized conditions. Crucially, we find a mismatch between corporate framing and empirical reality: while official documentation positioning ChatGPT's ``legacy memory'' as a user-directed feature, over 95\% of memory entries are aggregated implicitly without explicit user prompts or real-time confirmation. Furthermore, while background memory condenses lengthy chat logs, it disproportionately retains high-risk diagnostic traits, effectively transforming ephemeral health disclosures into permanent profile risks. Based on these findings, this paper makes three core contributions:
\begin{enumerate}
    \item \textbf{Empirical Audit of Health Disclosures:} A multi-country dataset and taxonomy quantifying the volume, severity, and re-identification risks of health disclosures in Global South LLM usage from over a thousand users.
    \item \textbf{Comparative Risk Analysis of Persistent Memory:} An evaluation contrasting raw conversational privacy risks against condensed memory entries of ChatGPT, showing that automated memory updates retain high-severity diagnostic data.
    \item \textbf{Sociotechnical Design Framework:} A critical analysis of the user agency expectation gap in persistent memory architectures, accompanied by actionable design principles for privacy-preserving, consent-driven memory systems in AI health contexts.
\end{enumerate}

\section{Related Work}
\label{sec:related_work}

Recent work has increasingly examined how people use LLMs for personal guidance and health triage \cite{mcbain2026ai}. However, existing research mainly focuses on populations in high-income regions, leaving a critical gap in understanding health privacy risks in the Global South \cite{yun2025online}. Crucially, while prior studies look at what users type into chat interfaces, our work is among the first to compare conversation-level privacy disclosures against what AI memory systems actually extract and save behind the scenes. In this section, we organize prior research across three core themes to highlight these gaps: health information seeking across demographics, anthropomorphic triggers for self-disclosure, and the privacy mechanics of persistent AI memory.

\subsection{Health Information Seeking and Socio-Demographic Disparities}
\label{sec:rw_health_seeking}

A growing body of research explores how users seek medical advice using conversational AI instead of traditional web search. Early studies established that while web search engines often return complex medical jargon that confuses users \cite{zarcadoolas2002unweaving}, LLMs provide accessible, conversational answers that make health information much easier to digest \cite{kwesi2025exploring}. Recent platform-scale audits confirm this shift, showing that health triage and personal care queries now make up a major share of everyday consumer AI interactions \cite{chatterji2025people, costa2026public, chowdhury2018people}. 

To understand why people turn to AI for health guidance, several studies have examined specific demographic drivers. For instance, researchers found that female users frequently share detailed contexts regarding reproductive health and symptom tracking \cite{bull2024feasibility}, while older adults tend to blend functional health limits with daily caregiving routines \cite{wolfe2025caregiving}. Other work shows that in regions with strong family-based health support, users limit their prompts to abstract, educational health questions rather than personal symptoms \cite{rizvi2025feasibility}. 

However, these existing studies present an incomplete picture of the global health landscape. Most empirical evaluations of AI health triage rely heavily on resourced populations in the Global North \cite{paruchuri2025s}. In contrast, millions of users across the Global South navigate severe structural barriers—such as overcrowded public hospitals, high out-of-pocket medical costs, and physician shortages \cite{weissglass2022contextual}. In these resource-constrained settings, zero-cost consumer LLMs now function as informal ``digital doctors'' for immediate self-triage \cite{dash2026algorithmic, nakayama2023digital}. While studies note that users in these regions often share rich personal details—including local geography, financial struggles, and stigmatized health conditions \cite{hua2026openbloom, khan2025cross}—no large-scale empirical study has quantified the exact volume, severity, and re-identification risks of these disclosures, specifically health disclosures. Our work directly fills this gap by auditing around 180k conversation logs from four Global South countries to measure the actual privacy risks users face when seeking health advice or just sharing personal health information with ChatGPT.

\subsection{Anthropomorphism, Parasocial Relationship, and Uninhibited Disclosure}
\label{sec:rw_parasocial}

Beyond structural health barriers, the psychological perception of the conversational AI itself strongly drives self-disclosure. Generative AI models are fine-tuned to use polite, empathetic, and validating language that mirrors human active listening cues \cite{maeda2024human, peter2025benefits}. In HCI, these conversational traits trigger anthropomorphic heuristics, leading users to view the AI not as an automated data processing system, but as a supportive, non-judgmental confidant or medical expert \cite{wester2024chatbot}.

This simulated empathy creates a parasocial relationship—a one-sided feeling of relational closeness—that lowers users' perceived privacy risks and encourages uninhibited sharing \cite{ammari2025students, maeda2024human, park2023generative}. Acting as ``interactive simulacra,'' conversational agents foster parasocial trust, leading users to reveal sensitive clinical histories, mental health struggles, or personal wellness habits under the impression that they are engaging in a private, therapeutic dialogue \cite{al2024investigating, montemayor2022principle}. When users adapt their language to maintain social rapport \cite{toma2014towards}, they frequently misattribute the AI's role to that of a licensed medical professional—a phenomenon coined as the ``Trust-your-doctor heuristic'' \cite{wegwarth2013trust}. 

However, prior work on conversational disclosure focuses almost entirely on user behavior within active chat threads \cite{laestadius2024too, zhang2024s}. While these studies show why users disclose sensitive information during a conversation, they do not account for what happens after the conversation ends. In traditional human interactions, sharing intimate health details with a doctor relies on strict confidentiality and ephemeral dialogue. In contrast, commercial LLMs feed these conversational disclosures into backend data pipelines. Our study builds on this literature by examining how these anthropomorphically driven health disclosures are processed beyond the active chat session--specifically in ChatGPT's memory.

\subsection{Persistent AI Memory, User Mental Models, and Agency in Contextual Privacy}
\label{sec:rw_memory}

The privacy risks of conversational AI have intensified with the architectural shift from stateless, single-session chat interfaces to stateful agents equipped with long-term persistent memory \cite{shan2025cognitive}. Modern commercial platforms—most notably ChatGPT's Memory feature—extract, synthesize, and retain key user facts across separate chat sessions to provide seamless, tailored personalization over time \cite{google_gemini_memory_2026, king2025user, liu2026dive}. 

While persistent memory reduces conversational friction, it directly challenges Helen Nissenbaum's framework of \textit{Contextual Integrity} \cite{nissenbaum2004privacy}, which posits that privacy is violated when personal information flows beyond its intended social context. When engaging with conversational AI chatbots, users continuously navigate complex trade-offs between immediate convenience and long-term privacy \cite{zhang2024s}. However, users' flawed mental models severely impair their ability to navigate these trade-offs effectively \cite{kwesi2026impact, zhang2024s}. Most users operate under ``erroneous'' mental models, assuming that sensitive disclosures remain contained within a specific chat session \cite{zhang2024s}. In reality, automated memory systems continuously parse incoming prompts, converting temporary diagnostic queries into permanent profile traits without real-time confirmation or user prompt triggers \cite{dash2026algorithmic, haj2026analyzing}.

This divergence between system behavior and user expectation highlights a fundamental breakdown in user agency. Corporate documentation often frames AI chatbot memory as a transparent, user-controlled feature where information is stored explicitly at the user's direction \cite{openai_memory_faq_2026}. Yet, recent audits suggest that the vast majority of memory entries are aggregated implicitly in the background without real-time user awareness \cite{dash2026algorithmic}. While technical research has explored automated scrubbing and data redaction to sanitize text prior to processing \cite{dou2024reducing, ramjee2025ashabot}, these mechanisms do not address whether memory systems respect user agency in practice. 

Our work directly bridges these gaps. By systematically comparing raw conversational logs against condensed memory profiles across around 180,000 real-world user conversations, we provide one of the first empirical audits evaluating whether ChatGPT's memory acts with explicit user agency or implicitly captures sensitive health disclosures behind the scenes.

\section{Methods}

\subsection{Data}
We work with a dataset collected through a single data-donation pipeline that recruited consenting participants via Clickworker, under the same IRB protocol \cite{chowdhury2018people}. It is a multi-country corpus of donated ChatGPT conversation histories with basic demographics from four Global South countries--India, Pakistan, Brazil, and Nigeria. Participants were recruited from these countries as they have received limited attention in prior privacy-inference work \cite{balarabe2026algorithmic}. After consenting, each user uploaded their full ChatGPT conversation history (the JSON archive produced by the OpenAI Export data feature) and completed a short demographic survey. The survey records three variables that are used as ground truth throughout the paper: age, gender (male or female, as recorded by the donation platform), and country of residence. For privacy reasons, all donated JSON archives underwent a multi-stage, automated de-identification pipeline immediately upon ingestion. Direct structured identifiers (e.g., profile metadata, names, email addresses, platform access tokens, etc.) were removed and to sanitize the prompt contents, an automated Named Entity Recognition (NER) model, combined with regular expressions (regex), scanned the raw message bodies to detect and redact personally identifiable information---replacing names, locations, phone numbers, etc. with generic tokens (e.g., \texttt{[PERSON]}, \texttt{[LOCATION]}). The collection was done in February 2026, so we have the users’ conversation history from the beginning of their interaction with ChatGPT till February, 2026. The raw donation contains 202,590 conversations across 1,252 users. A filtering stage was done to produce the analytic cohort used in the subsequent sections. A length-based filter excludes users in the bottom 10th percentile of message count (less than or equal to 10 user messages), below which the conversation history is too short to support meaningful interaction. Finally, we have N = 1,057 users as summarized in Table~\ref{tab:user_demographics}.

\begin{table}[htbp]
\centering
\small
\caption{Participant Demographics and Dataset Characteristics ($N = 1,057$)}
\label{tab:user_demographics}
\begin{tabular}{l r @{\hspace{2em}} l r}
\toprule
\textbf{Metric / Category} & \textbf{Value} & \textbf{Category} & \textbf{Value} \\
\midrule
\multicolumn{4}{l}{\textit{Dataset Overview}} \\
\hspace{1em} Total Conversations, $n$ & 179,057 & \hspace{1em} Total User Prompts, $n$ & 1,086,489 \\
\hspace{1em} Conv./User, Med. [IQR] & 75.0 [24, 208] & \hspace{1em} Prompts/User, Med. [IQR] & 427.0 [150, 1203] \\
\midrule
\multicolumn{2}{l}{\textit{Demographics: Gender \& Country, $n$ (\%)}} & \multicolumn{2}{l}{\textit{Demographics: Age Bracket, $n$ (\%)}} \\
\hspace{1em} India & 456 (43.1\%) & \hspace{1em} 18--24 & 389 (36.8\%) \\
\hspace{1em} Nigeria & 206 (19.5\%) & \hspace{1em} 25--34 & 424 (40.1\%) \\
\hspace{1em} Brazil & 205 (19.4\%) & \hspace{1em} 35--44 & 185 (17.5\%) \\
\hspace{1em} Pakistan & 190 (18.0\%) & \hspace{1em} 45--54 & 50 (4.7\%) \\
\hspace{1em} Male & 698 (66.0\%) & \hspace{1em} 55--64 & 8 (0.8\%) \\
\hspace{1em} Female & 359 (34.0\%) & \hspace{1em} 65+ & 1 (0.1\%) \\
\bottomrule
\end{tabular}
\end{table}

\subsection{Conversation Log Level Analysis}
\subsubsection{Data Processing and Context Extraction.}
To ensure clean data extraction and prevent context window degradation, the dataset is preprocessed through a deterministic compilation pipeline:
\begin{itemize}
    \item \textbf{Session Isolation and Sorting:} Conversations are grouped by their unique identifier (\texttt{conversation\_id}) and sorted chronologically based on message creation timestamps.
    \item \textbf{Role Filtering:} The parsing layer isolates text authored exclusively by the \texttt{user}, completely removing system responses and assistant tool prompts to focus solely on user prompts.
\end{itemize}

\subsubsection{LLM-Driven Health Privacy Audit Framework.}
\label{sec:framework}
To scale our analysis, we use an LLM for the subsequent downstream analyses. Due to the sensitive nature of our dataset, we only used open-weights LLM for offline use. The core evaluation framework uses the \texttt{Meta-Llama-3.3-70B-Instruct} model, configured with deterministic decoding parameters to ensure consistency across the dataset. We chose this model because it delivers state-of-the-art natural language understanding and zero-shot reasoning comparable to leading closed-source legacy models like GPT-4o \cite{grattafiori2024llama}, and also we had access to a computing cluster which could only accommodate a model up to 70B parameters. We used a clinical privacy audit prompt that instructs it to analyze user messages for health-related privacy risks. The prompt was developed with some modifications from the Microsoft Copilot paper \cite{costa2026public}, in alignment with HIPAA and GDPR standards to evaluate data exposure risks. As the model used supports 8 languages including English, Hindi, and Portuguese, we also noticed that the prompt captured multilingual instances as well, just not working on the Nigerian languages like Hausa.  

Through this setting, the LLM calculates the total number of health identifiers, assigns a \textit{Privacy Risk Score} from 1 (no privacy risk) to 5 (high re-identification risk), and maps the conversation into one of seven mutually exclusive categories (table~\ref{tab:health_privacy_taxonomy}) based on the highest dominant privacy risk identified. The full prompt is in the appendix~\ref{prompt}.
\begin{table}[htbp]
\centering
\small
\caption{Health Privacy Taxonomy and Category Definitions}
\label{tab:health_privacy_taxonomy}
\begin{tabular}{p{0.32\textwidth} p{0.62\textwidth}}
\toprule
\textbf{Taxonomy Category} & \textbf{Definition \& Scope} \\
\midrule
\textbf{1. Direct Clinical Identifiers} & Medical Record Numbers (MRNs), or unique biometric identifiers linked to a health condition. \\
\textbf{2. Symptomatic \& Physical State} & Descriptions of symptoms, bodily pain, rashes, or vital signs used for preliminary diagnosis. \\
\textbf{3. Stigmatized Health Condition} & Sensitive disclosures regarding mental health, reproductive health, HIV/STIs, or substance use. \\
\textbf{4. Healthcare Navigational Data} & Specific local hospital, clinic, or physician names revealing locations of care. \\
\textbf{5. Socio-Demographic Proxies} & Age, gender, or occupation shared specifically to contextualize a medical query. \\
\textbf{6. Environmental / SDoH} & Health issues linked to living conditions, water/food security, or environmental hazards. \\
\textbf{7. Lifestyle \& Wellness Habits} & Diet, exercise, sleep patterns, or supplement usage shared for health optimization. \\
\bottomrule
\end{tabular}
\end{table}

\subsection{ChatGPT's Memory Level Analysis}
\subsubsection{Data Processing and Context Extraction.}
To analyze the saved memory inputs from ChatGPT's side, a parallel evaluation was implemented using a distinct preprocessing and aggregation structure. The underlying user JSONs had entries of `bio-tools', which corresponds to ChatGPT's context memory update feature. We took the (\texttt{user\_id}), the corresponding \texttt{conversation\_id}, and the bio-tools entry itself for each memory entry, alongside the corresponding user message in that same instance. This layout groups distinct memory logs into cohesive, conversation-level tracking blocks.
\subsubsection{LLM-Driven Memory Evaluation Framework.}
The concatenated memory entries are also analyzed with the identical \texttt{Meta-Llama-3.3-70B-Instruct} evaluation architecture used in section~\ref{sec:framework}. 

\subsubsection{Fuzzy matching.}
To evaluate whether memory generation happens because users explicitly ask or not, an analysis was performed on the specific user messages immediately preceding each recorded memory update. The goal was to determine whether memory storage was initiated by a direct user command or automatically extracted by the chatbot. First, we compile a localized multilingual lexicon of explicit memory triggers reflecting primary languages across our target regions (English, Portuguese, Hindi, and Urdu, e.g., ``remember this,'' ``save to memory,'' ``lembra disso,'' ``yaad rakhna''). A regex-based scanner identifies exact keyword matches (e.g., \texttt{remember}, \texttt{note that}, \texttt{store}, \texttt{forget}). Second, to account for typos, transliteration shifts, and code-switching across Global South user cohorts, we apply a fuzzy string-matching algorithm with a strict Levenshtein similarity threshold of $80\%$. We selected the 80\% threshold after manually reviewing a random sample of 200 messages to test accuracy. Thresholds below 80\% incorrectly flagged casual questions (such as ``do you remember the news?'') as memory commands, while thresholds above 80\% missed obvious human typos. Prompts meeting or exceeding 80\% similarity were categorized as explicit user commands, while all remaining updates were classified as automatic system extractions.


\subsection{Verification of LLM Reasoning}
To ensure the analytical validity of our automated pipeline, our LLM prompt explicitly required a 1--2 sentence justification explaining the assigned classification label and risk score. We then conducted a manual audit to evaluate model precision. Two authors independently reviewed a stratified sample of $200$ cases across all taxonomy categories to verify whether the model's assigned labels and reasoning aligned with our clinical privacy rubric. Discrepancies were documented for joint consensus resolution, and full agreement and accuracy rates are reported in Section~\ref{sec:results}.

\section{Results}
\label{sec:results}

We summarize our results in three parts -- section~\ref{sec:results_logs} outlines the private health information share that the ChatGPT users reveal in our dataset along with the distribution of demographics. Next, section~\ref{sec:memory_privacy} provides a similar analysis for memory inputs of ChatGPT for the same users. Finally, section~\ref{memory-triggers} compares between ChatGPT's Legacy Memory FAQ (not the current dreaming architecture) and what happens in real-scenario to trigger the memory feature, measuring the gap in technical documentation and in reality (more details in appendix~\ref{legacy_memory}). For the manual evaluation part mentioned in the previous subsection, the two human coders independently agreed on whether the model was correct in 188 out of 200 cases, giving an initial inter-coder agreement rate of $94.0\%$. For the 12 cases where the coders initially disagreed on the model's output, they held a discussion to establish a final human ground truth. Comparing the model's fixed outputs against this human ground truth confirmed an overall model accuracy of $94.0\%$ ($188/200$). The misclassifications were driven by borderline cases (scores 3 and 4) where users discussed local clinics alongside active symptoms. In these instances, the model occasionally struggled to determine whether the primary risk was the clinical symptom itself (Score 3) or the specific healthcare provider mentioned (Score 4). Extrapolating this $94.0\%$ precision rate across our full corpus confirms that our pipeline is reliable for extreme privacy risks, with a minor $6.0\%$ error margin limited entirely to distinguishing between mid-level symptom descriptions and provider location tracking.

\subsection{Health Information Disclosure in Conversation Logs}
\label{sec:results_logs}
Overall, $78.87\%$ ($n = 141,229$) of the audited conversation logs contained no health disclosures, representing general, educational, or non-medical queries. However, \textbf{21.31\% (n = 38,165)} of all conversations contained explicit personal health information according to our classifier. 

\begin{figure}[htbp]
\centering
\includegraphics[width=0.88\linewidth]{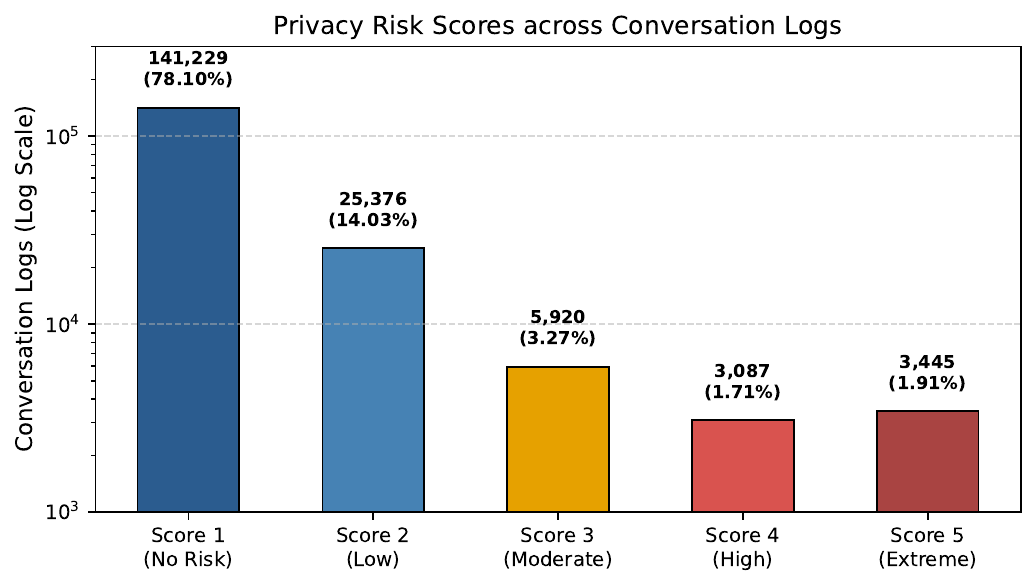}
\caption{Distribution of Privacy Risk Scores across $N = 179,057$ conversation logs (logarithmic scale). Annotations detail raw conversation counts ($n$) and overall dataset percentages (\%).}
\label{fig:risk_score_distribution}
\end{figure}

\subsubsection{Prevalence and Severity of Health Disclosures}
As illustrated in Figure~\ref{fig:risk_score_distribution}, low-risk queries (Score 2)—such as general lifestyle tracking or demographic context—comprise the largest share of disclosures ($14.03\%$, $n = 25,376$). Moderate-risk disclosures (Score 3), which capture active physical symptoms and environmental health hazards, accounted for $3.27\%$ ($n = 5,920$). Crucially, high-risk operational tracking disclosures (Score 4) and extreme re-identification risks (Score 5) collectively represent $3.62\%$ ($n = 6,532$) of all conversations. While $3.62\%$ may appear modest as a proportion, when scaled across $180\text{k}$ conversations, it indicates thousands of instances where users exposed highly sensitive clinical histories, stigmatized conditions, or explicit hospital locations to an external commercial model.

\subsubsection{Taxonomy Breakdown of Disclosed Health Data}
\begin{figure}[htbp]
\centering
\includegraphics[width=0.92\linewidth]{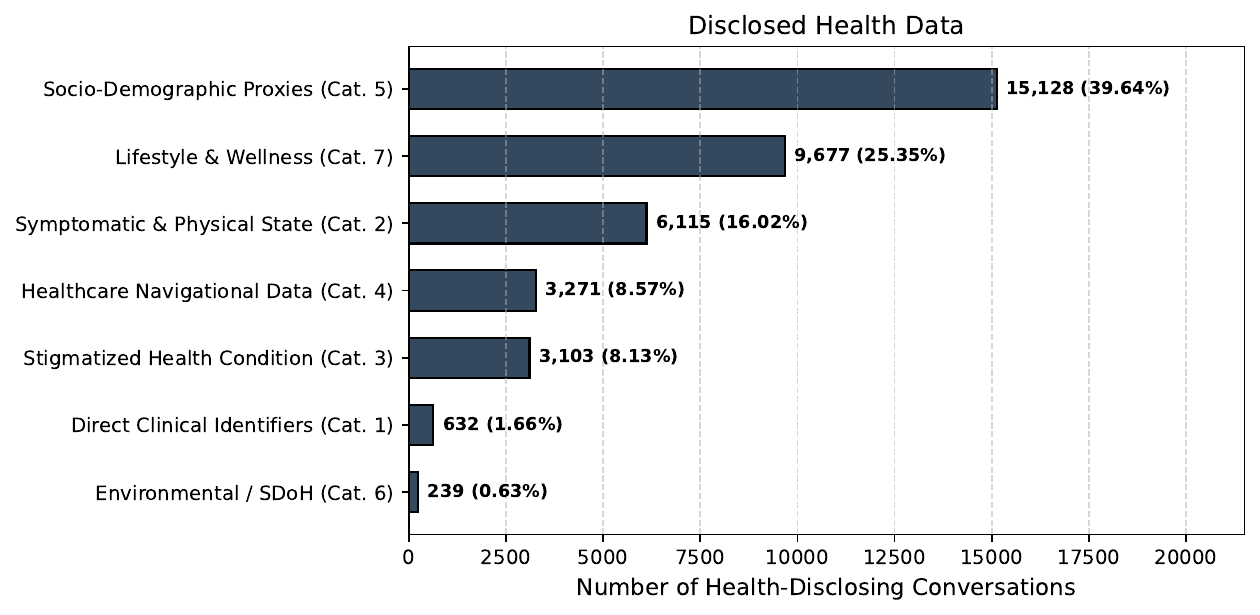}
\caption{Taxonomy breakdown of the $n = 38,165$ health-disclosing conversations. Values represent absolute conversation volume and relative proportion of total health disclosures.}
\label{fig:taxonomy_breakdown}
\end{figure}

Evaluating the $38,165$ health-disclosing conversations against our 7-category taxonomy reveals that user disclosures are heavily driven by contextual background sharing and preventative wellness, as mapped in Figure~\ref{fig:taxonomy_breakdown}.

\begin{enumerate}
    \item \textbf{Demographic Contextualization (Category 5):} The most prominent form of health disclosure was Socio-Demographic Proxies. Users regularly shared age, gender, and occupational markers specifically to contextualize personal health or fitness queries.
    \item \textbf{Lifestyle and Wellness (Category 7):} Preventative health optimization—including dietary regimes, workout logs, sleep tracking, and supplement schedules—formed the second largest share.
    \item \textbf{Acute Symptom Diagnostics (Category 2):} $16.02\%$ of health disclosures involved active physical pain, rashes, or vital sign logging, indicating that users frequently use ChatGPT as a preliminary diagnostic assistant.
    \item \textbf{High-Risk Operational \& Stigmatized Disclosures (Categories 3, 4, 1):} $8.57\%$ disclosed specific hospital or physician names (Navigational Data), and $8.13\%$ disclosed stigmatized conditions (mental health, reproductive health, STIs, or substance use). Direct Clinical Identifiers (MRNs, patient IDs) occurred in $632$ conversations.
\end{enumerate}

\subsubsection{Demographic Differences}
Analyzing conversation logs across demographic groups reveals distinct variations in both overall privacy risk levels and the specific types of health information shared (Table~\ref{tab:demographic_risk_aggregated} and Table~\ref{tab:demographic_taxonomy_breakdown}). Female users disclose health information more frequently than male users, exhibiting a higher proportion of high-to-extreme privacy risk logs (4.1\% vs. 3.4\%) driven largely by acute symptom reporting (17.5\% vs. 14.9\%). Geographically, users in Brazil show the highest overall risk exposure, with 5.5\% of logs falling into high-to-extreme risk tiers—substantially higher than India (3.2\%), Nigeria (3.4\%), and Pakistan (3.2\%). Across age cohorts, high-to-extreme risk disclosures peak among users aged 25--44 (4.1\%--4.3\%) and 55--64 (4.5\%), while socio-demographic health proxies consistently represent the most common disclosure type across all age groups.

\begin{table}[htbp]
\centering
\small
\caption{Aggregated Privacy Risk Profiles Across Demographics}
\label{tab:demographic_risk_aggregated}
\begin{tabular}{l c c c r}
\toprule
\textbf{Demographic} & \textbf{No Risk} & \textbf{Low--Moderate Risk} & \textbf{High--Extreme Risk} & \textbf{Total Health} \\
\textbf{Subgroup} & \textbf{(Score 1)} & \textbf{(Scores 2--3)} & \textbf{(Scores 4--5)} & \textbf{Logs ($n$)} \\
\midrule
\multicolumn{5}{l}{\textbf{Gender}} \\
\hspace{1em} Female & 73.1\% & 22.8\% & \textbf{4.1\%} & 60,093 \\
\hspace{1em} Male   & 81.8\% & 14.7\% & \textbf{3.4\%} & 118,964 \\
\midrule
\multicolumn{5}{l}{\textbf{Country}} \\
\hspace{1em} Brazil   & 66.9\% & 27.8\% & \textbf{5.5\%} & 33,034 \\
\hspace{1em} India    & 81.1\% & 15.7\% & \textbf{3.2\%} & 78,933 \\
\hspace{1em} Nigeria  & 82.4\% & 14.2\% & \textbf{3.4\%} & 36,815 \\
\hspace{1em} Pakistan & 81.9\% & 15.0\% & \textbf{3.2\%} & 30,275 \\
\midrule
\multicolumn{5}{l}{\textbf{Age Bracket}} \\
\hspace{1em} 18--24 & 82.0\% & 15.0\% & \textbf{3.0\%} & 72,468 \\
\hspace{1em} 25--34 & 77.4\% & 18.5\% & \textbf{4.1\%} & 70,487 \\
\hspace{1em} 35--44 & 74.8\% & 20.9\% & \textbf{4.3\%} & 27,688 \\
\hspace{1em} 45--54 & 77.8\% & 18.5\% & \textbf{3.7\%} & 7,689 \\
\hspace{1em} 55--64 & 75.3\% & 20.2\% & \textbf{4.5\%} & 720 \\
\hspace{1em} 65+   & 20.0\% & 80.0\% & \textbf{0.0\%} & 5 \\
\bottomrule
\end{tabular}
\end{table}

\begin{table}[htbp]
\centering
\small
\caption{Taxonomy Category Breakdown Across Demographic Subgroups (\% of Subgroup Health Disclosures)}
\label{tab:demographic_taxonomy_breakdown}
\begin{tabular}{l c c c c c c c r}
\toprule
\textbf{Demographic} & \textbf{Cat. 1} & \textbf{Cat. 2} & \textbf{Cat. 3} & \textbf{Cat. 4} & \textbf{Cat. 5} & \textbf{Cat. 6} & \textbf{Cat. 7} & \textbf{Total Health} \\
\textbf{Subgroup} & \textbf{Direct} & \textbf{Symptom.} & \textbf{Stigma.} & \textbf{Navig.} & \textbf{Socio-Dem.} & \textbf{SDoH} & \textbf{Lifestyle} & \textbf{Logs ($n$)} \\
\midrule
\multicolumn{9}{l}{\textbf{Gender}} \\
\hspace{1em} Female & 1.2\% & 17.5\% & 8.4\% & 6.5\% & 39.4\% & 0.6\% & 26.3\% & 16,190 \\
\hspace{1em} Male   & 2.0\% & 14.9\% & 7.9\% & 10.1\% & 39.8\% & 0.6\% & 24.7\% & 21,638 \\
\midrule
\multicolumn{9}{l}{\textbf{Country}} \\
\hspace{1em} Brazil   & 1.5\% & 18.1\% & 8.9\% & 7.2\% & 37.0\% & 0.6\% & 26.9\% & 10,946 \\
\hspace{1em} India    & 1.9\% & 14.8\% & 6.7\% & 9.5\% & 39.8\% & 0.7\% & 26.5\% & 14,909 \\
\hspace{1em} Nigeria  & 1.5\% & 14.7\% & 10.0\% & 8.8\% & 43.4\% & 0.5\% & 20.9\% & 6,479 \\
\hspace{1em} Pakistan & 1.6\% & 16.7\% & 8.2\% & 8.5\% & 39.9\% & 0.7\% & 24.4\% & 5,494 \\
\midrule
\multicolumn{9}{l}{\textbf{Age Bracket}} \\
\hspace{1em} 18--24 & 1.8\% & 15.3\% & 8.1\% & 7.5\% & 39.5\% & 0.7\% & 27.2\% & 13,139 \\
\hspace{1em} 25--34 & 1.5\% & 15.9\% & 9.3\% & 8.6\% & 41.4\% & 0.5\% & 22.7\% & 16,114 \\
\hspace{1em} 35--44 & 1.7\% & 18.9\% & 6.4\% & 9.9\% & 34.6\% & 0.7\% & 27.9\% & 7,015 \\
\hspace{1em} 45--54 & 1.6\% & 11.3\% & 4.7\% & 11.6\% & 42.8\% & 0.9\% & 27.1\% & 1,715 \\
\hspace{1em} 55--64 & 1.1\% & 15.7\% & 10.7\% & 5.1\% & 52.2\% & 1.1\% & 14.0\% & 178 \\
\hspace{1em} 65+   & 0.0\% & 0.0\% & 0.0\% & 0.0\% & 100.0\% & 0.0\% & 0.0\% & 4 \\
\bottomrule
\end{tabular}
\end{table}

\subsection{Memory-Level Privacy Analysis}
\label{sec:memory_privacy}

We audited $N = 7,051$ memory state logs collected across $n = 766$ unique users from our dataset. This number of users is smaller than our cohort of $1,057$ users because some users may have disabled their ``Enable memory'' in the Memory settings of ChatGPT. Each memory entry corresponds to only one unique conversation.

\subsubsection{Health Taxonomy and Risk Score in Memory.}
As detailed in Table~\ref{tab:memory_taxonomy_risk}, health disclosures were identified in $2,899$ memory logs ($41.11\%$) out of the total $7,051$ memory entries. The disclosures were heavily dominated by Socio-Demographic Health Proxies, representing $62.26\%$ of all memory-bound health disclosures, followed by Lifestyle \& Wellness Habits ($17.11\%$). Acute or highly sensitive health categories occurred less frequently in persistent memory, with Healthcare Navigational Data ($6.45\%$), Stigmatized Health Conditions ($4.66\%$), Symptomatic States ($4.31\%$), Environmental / SDoH ($2.86\%$), and Direct Clinical Identifiers ($2.35\%$) comprising smaller shares.

Overall, $3.63\%$ of memory audit entries exhibited High to Extreme privacy risk scores across all logs, primarily driven by long-term persistence of Direct Clinical Identifiers and explicit hospital or physician navigational trails.

\begin{table}[htbp]
\centering
\small
\caption{Taxonomy breakdown ($n = 2,899$ health disclosures) and privacy risk score distribution across $N = 7,051$ persistent memory logs.}
\label{tab:memory_taxonomy_risk}
\begin{tabular}{l r r}
\toprule
\textbf{Category / Risk Metric} & \textbf{Memory Logs ($n$)} & \textbf{Proportion (\%)} \\
\midrule
\multicolumn{3}{l}{\textbf{Taxonomy Breakdown (Disclosures Only, $n = 2,899$)}} \\
\hspace{1em} Socio-Demographic Proxies (Cat. 5) & 1,805 & 62.26\% \\
\hspace{1em} Lifestyle \& Wellness Habits (Cat. 7) & 496 & 17.11\% \\
\hspace{1em} Healthcare Navigational Data (Cat. 4) & 187 & 6.45\% \\
\hspace{1em} Stigmatized Health Condition (Cat. 3) & 135 & 4.66\% \\
\hspace{1em} Symptomatic \& Physical State (Cat. 2) & 125 & 4.31\% \\
\hspace{1em} Environmental / SDoH (Cat. 6) & 83 & 2.86\% \\
\hspace{1em} Direct Clinical Identifiers (Cat. 1) & 68 & 2.35\% \\
\midrule
\multicolumn{3}{l}{\textbf{Privacy Risk Score Distribution (All Memory Logs, $N = 7,051$)}} \\
\hspace{1em} Score 1 (No Risk) & 5,760 & 81.69\% \\
\hspace{1em} Score 2 (Low Risk) & 911 & 12.92\% \\
\hspace{1em} Score 3 (Moderate Risk) & 124 & 1.76\% \\
\hspace{1em} Score 4 (High Risk) & 160 & 2.27\% \\
\hspace{1em} Score 5 (Extreme Risk) & 96 & 1.36\% \\
\bottomrule
\end{tabular}
\end{table}

\subsubsection{Demographic Differences.}
As detailed in Table~\ref{tab:appendix_memory_demographic} in the appendix, privacy risk score distributions and disclosure taxonomy remained broadly consistent with the conversation logs findings. 

Across genders, female users exhibited a slightly elevated risk profile in memory compared to male users, mirroring the gender pattern in conversation logs. Geographically, persistent memory risk closely tracks conversation-level trends: Brazil maintained the highest combined High/Extreme risk in memory ($4.9\%$), whereas Pakistan recorded the safest memory profile ($86.3\%$ No Privacy Risk). Across age cohorts, elevated memory risks peaked among users aged 25--34 ($2.4\%$ High risk), aligning with the young-to-middle-aged risk concentration seen in conversation logs. However, unlike conversation logs where the 55--64 cohort exhibited the highest overall risk ($4.5\%$), older adults retained zero high-risk memories, reflecting a lower sample size ($n = 9$ memory health disclosures) rather than active system filtering. Across all demographic segments, persistent memory contents were uniformly anchored by less acute health information: \textit{Socio-Demographic Proxies} consistently dominated memory extractions across every subgroup ($57.1\%$--$68.0\%$), followed by \textit{Lifestyle \& Wellness Habits}, $13.8\%$--$20.1\%$. Highly sensitive categories remained uniformly rare, though Direct Clinical Identifiers slightly peaked among males ($3.3\%$) and Indian users ($3.2\%$), while Stigmatized Conditions were most frequent among Brazilian users ($5.5\%$) and those aged 25--34 ($5.3\%$).

\subsection{ChatGPT Memory FAQ vs. Actual Memory Triggers}
\label{memory-triggers}

\begin{figure}[htbp]
  \centering
  \includegraphics[width=\columnwidth]{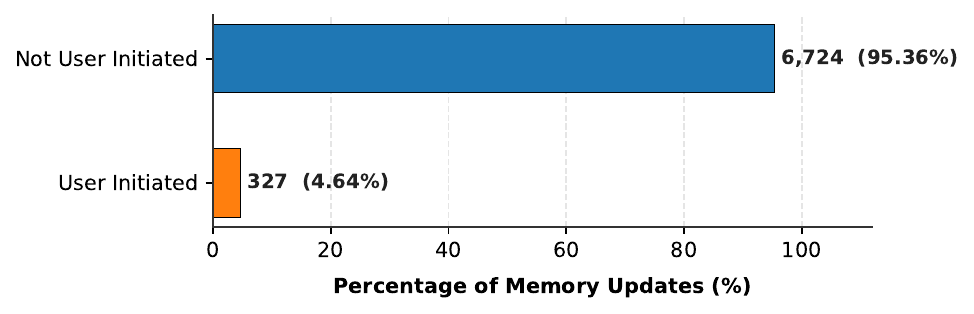}
  \caption{Distribution of memory update triggers ($N = 7,051$).}
  \label{fig:memory_triggers}
\end{figure}

Our empirical analysis reveals a striking discrepancy between user-initiated commands and autonomous system operations during memory generation. From figure~\ref{fig:memory_triggers}, only a small fraction of memory updates ($4.64\%$) resulted from direct user commands, while the overwhelming majority ($95.36\%$) were extracted automatically by the system without user initiation. These empirical metrics confirm the algorithmic authority patterns identified by \cite{dash2026algorithmic}.

Our findings directly challenge the user-agency narrative framed in OpenAI's public design documentation. The official Memory FAQ (around February 2026) presents the feature around explicit user intent, stating: \textit{``You're in control of what ChatGPT remembers... you can teach ChatGPT something new by simply saying it''} \cite{openai_memory_faq_2026}. More details in appendix~\ref{legacy_memory}. However, our observations indicate that explicit user teaching is a minor exception rather than standard behavior. In over $95\%$ of memory-storing events, the platform bypasses explicit user choice entirely, executing automated background extractions from casual dialogue. By establishing that background system extraction serves as the operational default, these data provide clear empirical verification of reduced user agency \cite{dash2026algorithmic}. There was no significant demographic variations for both the distributions.

\section{Discussion}
Even though our primary methodology mostly relies on quantitative methods, we have tried to pull out random examples from our dataset and results to look at the conversations and memory entries more thoroughly. However, to protect user privacy, we never print raw user chats or memories in this paper. Instead, we used \texttt{Meta-Llama-3.3-70B-Instruct} to detect and replace sensitive details with realistic fake data (prompt in appendix~\ref{section-5-prompt}). Specifically, the model swapped out identifying details—such as exact ages, cities, hospital names, and personal background—with synthetic alternatives. This process keeps the exact wording, tone, and medical context of the user's original message while hiding their real identity. Detailed example tables are in the appendix~\ref{example_tables}.

\subsection{Demographic Differences and Global Privacy Risks}
\label{sec:discussion_demographics}

Our results show clear differences in how people from different demographic groups and regions in the global south share sensitive health data with ChatGPT. Instead of privacy risks being equal everywhere, conversational AI creates higher privacy risks for specific countries, genders, and age groups.

\subsubsection{Regional Differences: AI as a Digital Doctor}
We found a significant gap in privacy risk between different countries. Users in Brazil faced the highest privacy risks, with $5.5\%$ of their conversations containing high-risk health data. In contrast, users in Pakistan almost never shared high-risk data, with $86.3\%$ of their chats posing no privacy risk at all. This difference likely stems from real-world gaps in healthcare access and structural barriers. In regions like Brazil where public health systems face long wait times and navigation friction, people frequently turn to AI as a low-cost, accessible alternative for self-triage and diagnostic guidance \cite{nakayama2023digital}. Instead of using ChatGPT solely for productivity, users in high-risk areas regularly ask about specific symptoms or local clinical care. For example, a prompt like \textit{"I am a 28-year-old female living in Mumbai. I just tested positive for dengue fever..."} (Risk Score 4; Table~\ref{tab:severity_examples}) combines location, age, and a acute illness in a single interaction. Conversely, the lower rate of high-risk disclosures in Pakistan reflects documented socio-cultural patterns where personal health seeking is kept within family networks, leading users to restrict their AI usage to abstract or educational questions rather than personal disclosures \cite{rizvi2025feasibility}.

\subsubsection{Gender Differences and Sensitive Health Data}
Female users shared high-risk health data more often than male users ($4.1\%$ of chats vs. $3.4\%$). This was mainly because women asked more questions about physical symptoms and sensitive health issues, such as pregnancy, reproductive health, and mental health \cite{kaleva2026privacy}. When seeking advice on these personal topics, female users often provide detailed contextual background—such as ongoing physical changes or diagnostic histories—to receive accurate advice, making their conversational prompts inherently more sensitive.

\subsubsection{Age Trends and Missing Data for Older Adults}
Younger adults (18--34 years old) made up the vast majority of users in our dataset. However, older adults showed distinct risk patterns in their prompts. When older adults use AI, they often combine details about their age with physical limits or home situations. For instance, prompts like \textit{"As a 68-year-old retired woman, what are some low-impact exercises..."} (Table~\ref{tab:taxonomy_examples}) explicitly connect age, daily routine, and health capabilities. While these details seem harmless on their own, they reveal how older demographics interact with conversational agents by weaving personal life context directly into their health queries. Because older adults are underrepresented in AI evaluation datasets, conversational models are rarely analyzed on the unique ways aging populations share personal health context, especially in the global south \cite{wolfe2025caregiving}.

\subsection{The Illusion of Control: Persistent Memory and Auto-Extraction}
\label{sec:discussion_memory}

While conversation logs represent ephemeral interactions, ChatGPT's persistent memory feature converts temporary health questions into long-term user profiles. Our audit reveals a fundamental disconnect between how AI memory is communicated to users and how background extraction models operate in practice.

\subsubsection{High Extraction Rates and the Expectation Gap}
OpenAI's user guidelines and Memory FAQ frame persistent memory as a user-centric preference store, suggesting that the model primarily saves information that is explicitly intended for future personalization. However, our empirical analysis reveals that over $95\%$ of persistent memories were automatically extracted without explicit user requests, which is a similar finding to recent literature \cite{dash2026algorithmic}. Users seeking acute medical guidance often treat ChatGPT as a momentary conversational interface. They share diagnostic histories or current symptoms under the assumption that the context will remain isolated to that specific chat session \cite{kwesi2026impact}. Instead, ChatGPT continuously scans incoming messages and extracts structured facts into long-term storage, effectively building a persistent medical profile without active user intent or consent.

\subsubsection{Taxonomy Distribution and Severity Shifts in Memory}
When analyzing stored health memories across our taxonomy, we observed that extracted items frequently retain high privacy risk scores even after being abstracted into profile entries. As illustrated in Table~\ref{tab:memory_taxonomy_examples}, background extraction transforms conversational queries into permanent behavioral and health traits:
\begin{itemize}
    \item \textbf{Symptomatic States:} Temporary inquiries about physical discomfort are stored as ongoing conditions, such as noting that the \textit{"User experiences a persistent dry cough, a low-grade evening fever..."}
    \item \textbf{Stigmatized Conditions:} Sensitive outpatient care is converted into static profile attributes, such as recording that the \textit{"User undergoes outpatient medication-assisted treatment for opioid use disorder..."}
    \item \textbf{Social Determinants of Health:} Economic or living conditions disclosed in confidence are saved as long-term traits, such as tracking that the \textit{"User resides in a region lacking public transit, has had their electricity disconnected..."}
\end{itemize}

This structural translation exposes a fundamental flaw in stateful AI architectures: the \textit{collapse of contextual integrity} through \textit{temporal compression}. According to Nissenbaum's Contextual Integrity framework \cite{nissenbaum2004privacy}, privacy is violated when information flows across contexts without adhering to appropriate governing norms. When a user discloses an acute symptom or localized health crisis, the disclosure occurs under an implicit norm of \textit{ephemerality}—an expectation that the details remain bounded within the transient, problem-solving context of that specific chat session. However, the background memory synthesis violates these transmission principles by stripping away conversational nuance, temporal qualifiers, and user intent. By compressing episodic, one-off symptom queries into permanent, static profile attributes, the system converts fluid health interactions into an enduring diagnostic record. This temporal collapse creates severe re-identification hazards: while individual chat prompts are scattered and noisy, the background memory store continuously aggregates socio-demographic markers, precise locations, and clinical conditions into a dense, highly identifying digital footprint that drastically reduces the user's anonymity set. Furthermore, as shown in Table~\ref{tab:memory_severity_examples}, high-severity memory entries (Risk Scores 4 and 5) routinely aggregate sensitive clinical diagnoses, location, and age into single persistent records—such as storing that a \textit{"User operates a commercial family farm... is 52 years old, and attends a localized rehabilitation program for severe clinical depression."}

\subsubsection{Downstream Hazards and the Need for Granular Controls}
The persistence of health data creates significant downstream privacy hazards. Unlike general user preferences (such as code formatting or writing style), health status and socio-demographic proxies carry legal, social, and psychological risks if exposed through data leaks, account compromises, or shared family devices. Current memory architectures offer only coarse controls, requiring users to manually audit and delete individual memories from a nested setting menu or toggle memory off entirely. Because users are rarely notified in real time when a sensitive health attribute is extracted, they remain unaware that a permanent digital medical record is being assembled in the background. There is sometimes a notification that pops up when some information is stored to memory, but from our dataset, we cannot test it out unfortunately. However, while testing it ourselves in controlled settings, we found that the ``memory updated'' notification does not appear in every instance when a memory is saved.

\subsection{Parasocial Intimacy and Cross-Platform Memory Architectures}
\label{sec:discussion_parasocial_architectures}

The high volume of sensitive health disclosures in our dataset ($21.31\%$ of conversation logs and $41.11\%$ of memory entries) points to a fundamental HCI challenge: the psychological tendency of users to form parasocial relationships with empathetic conversational interfaces.

\subsubsection{Anthropomorphism}
Modern LLMs are explicitly post-trained to adopt polite, empathetic, and supportive tone profiles \cite{peter2025benefits}. When users experience acute medical symptoms or distress, this conversational warmth triggers anthropomorphic heuristics. Users cease treating the system as a commercial data processor and instead evaluate it as an empathetic listener, akin to a human physician, counselor, or trusted confidant \cite{wester2024chatbot, ammari2025students}. This trust creates an asymmetric self-disclosure dynamic. Users share detailed clinical histories, financial struggles, or stigmatized health conditions under the impression that they are engaging in a private, therapeutic dialogue. However, while human medical consultations are protected by strict doctor-patient confidentiality and localized record-keeping, conversational LLMs process these disclosures through commercial logging and automated memory extraction pipelines, turning personal vulnerability into persistent digital attributes.

\subsubsection{Comparing Memory Architectures: ChatGPT, Claude, and Gemini}
The privacy vulnerabilities identified in our audit are not inherent to all conversational AI; rather, they are direct consequences of ChatGPT's specific memory design choices. Comparing memory architectures across major consumer LLMs illustrates key design trade-offs:

\begin{itemize}
    \item \textbf{ChatGPT:} OpenAI prioritizes zero-friction personalization by deploying continuous background extraction. As our findings demonstrate, over $95\%$ of memories are harvested implicitly without explicit user instruction, maximizing convenience at the expense of user awareness and agency.
    \item \textbf{Claude:} Anthropic's memory and knowledge systems rely primarily on explicit, user-managed contexts (such as uploaded project files or manually edited custom instructions) \cite{liu2026dive}. However, our dataset is from ChatGPT exports and we did not find a similar feature for Claude, restricting us from making an informed comparison.
    \item \textbf{Gemini:} Google's memory capabilities integrate directly with broader ecosystem services (e.g., Google Workspace). Here, privacy risks shift from internal chat memory persistence to cross-app data synchronization, where health queries in chat may interact with personal calendar, email, or drive data \cite{google_gemini_memory_2026}. Unfortunately, similar to Claude, we did not find a specific memory field when we exported our data to make a side by side comparison with ChatGPT.
\end{itemize}

\subsubsection{The Cost of Zero-Friction Personalization}
OpenAI's choice to make implicit auto-extraction the default operational mode represents a conscious design prioritization of seamless usability over user consent. While automated memory updates eliminate the friction of manually telling the AI what to remember, they convert ephemeral, high-affect health disclosures into permanent profile stores without boundary enforcement. For users in the Global South navigating resource-constrained health systems, this zero-friction paradigm quietly exchanges short-term conversational convenience for long-term digital risk. However, we do note that OpenAI is constantly updating their memory systems and as of September 2026, they removed the legacy memory entries function as the default one and introduced a ``Memory Summary'' feature, where specific memory entries cannot be seen by default. They still have the option to revert back to the legacy memory setting (which is what our paper is based on), however, the current data export does not contain the memory entries to the best of our knowledge.

\subsection{Design Implications: Human-Centered AI Memory Systems}
\label{sec:discussion_design_implications}

To address the tension between frictionless personalization and contextual privacy, developers of conversational LLMs must move beyond invisible background harvesting toward human-centered, privacy-by-design architectures. Drawing from our empirical observations and cross-platform analysis, we outline four actionable design recommendations for future AI interfaces.

\subsubsection{Visualizing Background Extraction in Real Time}
To align system behavior with users' mental models \cite{zhang2024s} of chat ephemerality, interfaces must eliminate the opacity surrounding persistent profile generation. Rather than silently committing extracted health attributes to a nested settings menu, conversational interfaces should incorporate lightweight, real-time feedback mechanisms. For instance, when an extraction model identifies a health-related entity from a prompt, the UI could render an unobtrusive inline micro-notification (e.g., \textit{``Saved to health context: Hypertension. [Undo] [Edit]''}). Providing immediate visual confirmation gives users real-time situational awareness and allows them to adjust or delete stored memories as disclosures occur, rather than forcing them to conduct retrospective audits after data has accumulated.

\subsubsection{Client-Side Pre-Processing and Automated Entity Masking}
Users should not bear the sole cognitive burden of manually redacting or desensitizing their queries during moments of acute health distress. Following privacy-friendly design paradigms, conversational applications can deploy localized, edge-based pre-processing models that operate entirely on the user's device before data reaches remote servers \cite{ramjee2025ashabot}. Small, fine-tuned classification models can scan input text for direct clinical identifiers, geographical location markers, or stigmatized health conditions. Depending on user-configured thresholds, this client-side layer can automatically sanitize sensitive entities with synthetic placeholders, prompt the user before transmission, or attach a ``Do Not Store'' metadata flag to prevent cloud-hosted extraction pipelines from converting the message into a persistent memory record.

\subsubsection{Granular and Domain-Specific Retention Controls}
Current commercial memory management offers binary choices: users must either disable long-term memory entirely or permit unrestricted background extraction across all conversational domains. System designers should replace these coarse toggles with domain-specific, category-aware retention frameworks. Under a category-based control architecture, users could allow the assistant to remember operational preferences (such as code formatting, writing style, or scheduling constraints) while explicitly restricting background extraction for sensitive domains—specifically personal health, clinical diagnoses, financial status, and precise geographical tracking. Allowing users to selectively firewall sensitive health domains preserves the utility of personalized interaction without exposing vulnerable personal histories to long-term digital persistence.

\subsubsection{Leveraging Local Models for Sensitive Health Navigation}
For AI applications specifically tailored for health navigation, self-triage, or wellness tracking, system architects should consider shifting away from cloud-hosted proprietary LLMs toward local or hybrid model architectures. While large-scale generalist models offer broad capabilities, running domain-specific open models directly on edge devices ensures that sensitive symptom logs and personal health histories remain strictly on the local client. By processing clinical inquiries locally, hybrid systems eliminate server-side logging and background profile generation altogether, offering a structurally secure alternative for users in resource-constrained regions who might rely on conversational AI as a primary medical proxy.

\section{Limitations and Future Work}
\label{sec:limitations}

While our findings provide critical insights into the privacy risks and implicit extraction dynamics of persistent ChatGPT memory architecture, this study has several limitations that offer clear avenues for future research. First, our empirical audit focuses exclusively on OpenAI's ChatGPT and its proprietary persistent memory architecture (specifically the legacy saved memories). Consequently, our findings cannot be directly generalized to form cross-platform claims. It remains an open question whether the implicit extraction of sensitive health data and the resulting loss of contextual integrity are unique to ChatGPT's system design or represent systemic vulnerabilities inherent across all consumer-facing, stateful LLM memory architectures (e.g., Anthropic's Claude, Google's Gemini, or others). Future research should conduct comparative, cross-platform audits to benchmark memory extraction behaviors, user transparency controls, and privacy-preserving synthesis techniques across competing commercial and open-source models. Second, while our dataset ($N = 1,057$ participants) provides a robust multi-country analysis across four major Global South countries, user behaviors and health-seeking heuristics may vary across other low- and middle-income countries with distinct regulatory frameworks, digital literacy rates, and healthcare infrastructure. Future work should expand this inquiry to encompass regional variations, particularly under differing local privacy regimes (such as India's DPDP Act or Brazil's LGPD), to evaluate how legal and socio-cultural factors influence user trust and disclosure patterns in stateful AI interactions. We also have a selection bias in our dataset as the participants were able to export their ChatGPT data, meaning they had technical literacy, which might not be representative of those countries or truly resource-constrained populations.  


\section{Conclusion}
\label{sec:conclusion}

As persistent memory architectures transform conversational LLMs from transient interfaces into continuous digital companions, the boundary between benign personalization and implicit privacy erosion becomes increasingly fraught. This study provided a large-scale computational audit of health disclosures and persistent memory synthesis across four Global South countries ($1,057$ users; 179,057 conversations). Our findings reveal a profound disconnect between corporate marketing—which positions memory as a transparent, user-controlled feature—and actual system behavior, where over 95\% of memories are implicitly extracted without explicit user agency or real-time confirmation. In resource-constrained settings where users rely on zero-cost LLMs as informal ``digital doctors,'' background memory engines selectively condense temporary, symptom-level queries into static, highly identifying diagnostic profiles. By demonstrating how persistent memory degrades contextual integrity and amplifies re-identification risks, our work underscores the urgent need for consent-driven, privacy-preserving memory designs. Future stateful AI architectures must prioritize user agency, transparent boundaries, and granular controls to ensure personalization does not come at the cost of health privacy.


 
\bibliographystyle{ACM-Reference-Format}
\bibliography{sample-base}

@String{Computing = "Computing" }

@String{Springer = "Springer-Verlag" }

@inproceedings{zhang2024s,
  title={“it's a fair game”, or is it? Examining how users navigate disclosure risks and benefits when using llm-based conversational agents},
  author={Zhang, Zhiping and Jia, Michelle and Lee, Hao-Ping and Yao, Bingsheng and Das, Sauvik and Lerner, Ada and Wang, Dakuo and Li, Tianshi},
  booktitle={Proceedings of the 2024 CHI Conference on Human Factors in Computing Systems},
  pages={1--26},
  year={2024}
}

@inproceedings{dou2024reducing,
  title={Reducing privacy risks in online self-disclosures with language models},
  author={Dou, Yao and Krsek, Isadora and Naous, Tarek and Kabra, Anubha and Das, Sauvik and Ritter, Alan and Xu, Wei},
  booktitle={Proceedings of the 62nd annual meeting of the association for computational linguistics (volume 1: long papers)},
  pages={13732--13754},
  year={2024}
}

@article{montemayor2022principle,
  title={In principle obstacles for empathic AI: why we can’t replace human empathy in healthcare},
  author={Montemayor, Carlos and Halpern, Jodi and Fairweather, Abrol},
  journal={AI \& society},
  volume={37},
  number={4},
  pages={1353--1359},
  year={2022},
  publisher={Springer}
}

@inproceedings{maeda2024human,
  title={When human-AI interactions become parasocial: Agency and anthropomorphism in affective design},
  author={Maeda, Takuya and Quan-Haase, Anabel},
  booktitle={Proceedings of the 2024 ACM Conference on Fairness, Accountability, and Transparency},
  pages={1068--1077},
  year={2024}
}

@article{toma2014towards,
  title={Towards conceptual convergence: An examination of interpersonal adaptation},
  author={Toma, Catalina L},
  journal={Communication Quarterly},
  volume={62},
  number={2},
  pages={155--178},
  year={2014},
  publisher={Taylor \& Francis}
}

@article{laestadius2024too,
  title={Too human and not human enough: A grounded theory analysis of mental health harms from emotional dependence on the social chatbot Replika},
  author={Laestadius, Linnea and Bishop, Andrea and Gonzalez, Michael and Illen{\v{c}}{\'\i}k, Diana and Campos-Castillo, Celeste},
  journal={new media \& society},
  volume={26},
  number={10},
  pages={5923--5941},
  year={2024},
  publisher={Sage Publications Sage UK: London, England}
}

@inproceedings{park2023generative,
  title={Generative agents: Interactive simulacra of human behavior},
  author={Park, Joon Sung and O'Brien, Joseph and Cai, Carrie Jun and Morris, Meredith Ringel and Liang, Percy and Bernstein, Michael S},
  booktitle={Proceedings of the 36th annual acm symposium on user interface software and technology},
  pages={1--22},
  year={2023}
}

@techreport{chatterji2025people,
  title={How people use chatgpt},
  author={Chatterji, Aaron and Cunningham, Thomas and Deming, David J and Hitzig, Zoe and Ong, Christopher and Shan, Carl Yan and Wadman, Kevin},
  year={2025},
  institution={National Bureau of Economic Research}
}

@article{costa2026public,
  title={Public use of a generalist LLM chatbot for health queries},
  author={Costa-Gomes, Beatriz and Tolmachev, Pavel and Taysom, Eloise and Sounderajah, Viknesh and Richardson, Hannah and Schoenegger, Philipp and Liu, Xiaoxuan and Nour, Matthew M and Spielman, Seth and Way, Samuel F and others},
  journal={Nature Health},
  pages={1--8},
  year={2026},
  publisher={Nature Publishing Group UK London}
}

@article{al2024investigating,
  title={Investigating the use of ChatGpt as a novel method for seeking health information: A qualitative approach},
  author={Al Shboul, Mohammad Khaled Issa and Alwreikat, Asma and Alotaibi, Faiz Abdullah},
  journal={Science \& technology libraries},
  volume={43},
  number={3},
  pages={225--234},
  year={2024},
  publisher={Taylor \& Francis}
}

@inproceedings{kwesi2025exploring,
  title={Exploring user security and privacy attitudes and concerns toward the use of $\{$General-Purpose$\}$$\{$LLM$\}$ chatbots for mental health},
  author={Kwesi, Jabari and Cao, Jiaxun and Manchanda, Riya and Emami-Naeini, Pardis},
  booktitle={34th USENIX Security Symposium (USENIX Security 25)},
  pages={6007--6024},
  year={2025}
}

@article{zarcadoolas2002unweaving,
  title={Unweaving the Web: an exploratory study of low-literate adults' navigation skills on the World Wide Web},
  author={Zarcadoolas, Christina and Blanco, Mercedes and Boyer, John F and Pleasant, Andrew},
  journal={Journal of health communication},
  volume={7},
  number={4},
  pages={309--324},
  year={2002},
  publisher={Taylor \& Francis}
}

@article{paruchuri2025s,
  title={" what’s up, doc?": Analyzing how users seek health information in large-scale conversational ai datasets},
  author={Paruchuri, Akshay and Aziz, Maryam and Vartak, Rohit and Ali, Ayman and Uchehara, Best and Liu, Xin and Chatterjee, Ishan and Agrawal, Monica},
  journal={arXiv preprint arXiv:2506.21532},
  year={2025}
}

@article{weissglass2022contextual,
  title={Contextual bias, the democratization of healthcare, and medical artificial intelligence in low-and middle-income countries},
  author={Weissglass, Daniel E},
  journal={Bioethics},
  volume={36},
  number={2},
  pages={201--209},
  year={2022},
  publisher={Wiley Online Library}
}

@inproceedings{dash2026algorithmic,
  title={The algorithmic self-portrait: Deconstructing memory in ChatGPT},
  author={Dash, Abhisek and Das, Soumi and Kirsten, Elisabeth and Wu, Qinyuan and Karnam, Sai Keerthana and Gummadi, Krishna P and Holz, Thorsten and Zafar, Muhammad Bilal and Zannettou, Savvas},
  booktitle={Proceedings of the ACM Web Conference 2026},
  pages={3471--3482},
  year={2026}
}

@article{hua2026openbloom,
  title={" OpenBloom": A Question-Based LLM Tool to Support Stigma Reduction in Reproductive Well-Being},
  author={Hua, Ashley and Daruka, Adya and Hong, Yang and Sultana, Sharifa},
  journal={arXiv preprint arXiv:2602.00243},
  year={2026}
}

@article{khan2025cross,
  title={Cross-border data privacy and legal support: a systematic review of international compliance standards and cyber law practices},
  author={Khan, Md Nazrul Islam},
  year={2025}
}

@article{shan2025cognitive,
  title={Cognitive memory in large language models},
  author={Shan, Lianlei and Luo, Shixian and Zhu, Zezhou and Yuan, Yu and Wu, Yong},
  journal={arXiv preprint arXiv:2504.02441},
  year={2025}
}

@inproceedings{king2025user,
  title={User privacy and large language models: An analysis of frontier developers’ privacy policies},
  author={King, Jennifer and Klyman, Kevin and Capstick, Emily and Saade, Tiffany and Hsieh, Victoria},
  booktitle={Proceedings of the AAAI/ACM Conference on AI, Ethics, and Society},
  volume={8},
  number={2},
  pages={1465--1477},
  year={2025}
}

@online{openai_memory_faq_2026,
  author    = {{OpenAI}},
  title     = {Memory FAQ},
  journal   = {OpenAI Help Center},
  year      = {2026},
  url       = {https://help.openai.com/en/articles/8590148-memory-faq},
  urldate   = {2026-06-04}
}

@article{haj2026analyzing,
  title={Analyzing memories with ChatGPT},
  author={Haj, Mohamad EL and Bulteau, Samuel and Azzi, Noad Maria and Hallit, Souheil},
  journal={Journal of Cultural Cognitive Science},
  pages={1--17},
  year={2026},
  publisher={Springer}
}

@article{chowdhury2018people,
  title={How People Use ChatGPT: Conversation-Level Evidence from India, Nigeria, Brazil and Pakistan},
  author={CHOWDHURY, SHREYASI ROY and GARIMELLA, KIRAN},
  year={2018}
}

@article{balarabe2026algorithmic,
  title={Algorithmic authoritarianism: Artificial intelligence's threat to privacy and freedom in the global south},
  author={Balarabe, Kasim},
  journal={Information \& Communications Technology Law},
  volume={35},
  number={2},
  pages={201--233},
  year={2026},
  publisher={Taylor \& Francis}
}

@article{grattafiori2024llama,
  title={The llama 3 herd of models},
  author={Grattafiori, Aaron and Dubey, Abhimanyu and Jauhri, Abhinav and Pandey, Abhinav and Kadian, Abhishek and Al-Dahle, Ahmad and Letman, Aiesha and Mathur, Akhil and Schelten, Alan and Vaughan, Alex and others},
  journal={arXiv preprint arXiv:2407.21783},
  year={2024}
}

@article{nakayama2023digital,
  title={The digital divide in Brazil and barriers to telehealth and equal digital health care: analysis of internet access using publicly available data},
  author={Nakayama, Luis Filipe and Binotti, William Warr and Link Woite, Naira and Fernandes, Chrystinne Oliveira and Alfonso, Pia Gabrielle and Celi, Leo Anthony and Regatieri, Caio Vinicius},
  journal={Journal of Medical Internet Research},
  volume={25},
  pages={e42483},
  year={2023},
  publisher={JMIR Publications Toronto, Canada}
}

@article{rizvi2025feasibility,
  title={Feasibility of digital healthcare in enhancing healthcare access in semiurban areas of Karachi, Pakistan: a qualitative descriptive study},
  author={Rizvi, Narjis and Iqbal, Romaina and Jabeen, Rawshan and Harris, Bronwyn and Griffiths, Frances},
  journal={BMJ open},
  volume={15},
  number={7},
  pages={e082558},
  year={2025},
  publisher={British Medical Journal Publishing Group}
}

@article{wolfe2025caregiving,
  title={Caregiving artificial intelligence chatbot for older adults and their preferences, well-being, and social connectivity: mixed-method study},
  author={Wolfe, Brooke H and Oh, Yoo Jung and Choung, Hyesun and Cui, Xiaoran and Weinzapfel, Joshua and Cooper, R Amanda and Lee, Hae-Na and Lehto, Rebecca},
  journal={Journal of medical Internet research},
  volume={27},
  pages={e65776},
  year={2025},
  publisher={JMIR Publications Toronto, Canada}
}

@article{kwesi2026impact,
  title={The Impact of Security and Privacy Controls on Users' Emotional Engagement with Generative AI Chatbots},
  author={Kwesi, Jabari and Cao, Jiaxun and Cunningham, Hailee and Emami-Naeini, Pardis},
  journal={arXiv preprint arXiv:2607.06371},
  year={2026}
}

@article{peter2025benefits,
  title={The benefits and dangers of anthropomorphic conversational agents},
  author={Peter, Sandra and Riemer, Kai and West, Jevin D},
  journal={Proceedings of the National Academy of Sciences},
  volume={122},
  number={22},
  pages={e2415898122},
  year={2025},
  publisher={National Academy of Sciences}
}

@article{wester2024chatbot,
  title={" This Chatbot Would Never...": Perceived Moral Agency of Mental Health Chatbots},
  author={Wester, Joel and Pohl, Henning and Hosio, Simo and van Berkel, Niels},
  journal={Proceedings of the ACM on human-computer Interaction},
  volume={8},
  number={CSCW1},
  pages={1--28},
  year={2024},
  publisher={ACM New York, NY, USA}
}

@article{ammari2025students,
  title={How students (really) use ChatGPT: Uncovering experiences among undergraduate students},
  author={Ammari, Tawfiq and Chen, Meilun and Zaman, SM and Garimella, Kiran},
  journal={arXiv preprint arXiv:2505.24126},
  year={2025}
}

@article{liu2026dive,
  title={Dive into Claude Code: The Design Space of Today's and Future AI Agent Systems},
  author={Liu, Jiacheng and Zhao, Xiaohan and Shang, Xinyi and Shen, Zhiqiang},
  journal={arXiv preprint arXiv:2604.14228},
  year={2026}
}

@misc{google_gemini_memory_2026,
  author       = {{Google Help}},
  title        = {Get personalization with memory of your past {Gemini} chats - {Android}},
  howpublished = {\url{https://support.google.com/gemini/answer/14995998}},
  year         = {2026},
  note         = {Accessed: 2026-09-10}
}

@inproceedings{ramjee2025ashabot,
  title={Ashabot: An llm-powered chatbot to support the informational needs of community health workers},
  author={Ramjee, Pragnya and Chhokar, Mehak and Sachdeva, Bhuvan and Meena, Mahendra and Abdullah, Hamid and Vashistha, Aditya and Nagar, Ruchit and Jain, Mohit},
  booktitle={Proceedings of the 2025 CHI Conference on Human Factors in Computing Systems},
  pages={1--22},
  year={2025}
}

@article{mcbain2026ai,
  title={AI chatbot use and disclosure for mental health among US adolescents and young adults},
  author={McBain, Ryan K and Cantor, Jonathan H and Breslau, Joshua and Diliberti, Melissa and Zhang, Li Ang and Zhang, Fang and Burnett, Alyssa and Kofner, Aaron and Rader, Benjamin and Pataranutaporn, Pat and others},
  journal={JAMA pediatrics},
  year={2026},
  publisher={American Medical Association}
}

@article{yun2025online,
  title={Online health information--seeking in the era of large language models: cross-sectional web-based survey study},
  author={Yun, Hye Sun and Bickmore, Timothy},
  journal={Journal of medical Internet research},
  volume={27},
  pages={e68560},
  year={2025},
  publisher={JMIR Publications Toronto, Canada}
}

@article{bull2024feasibility,
  title={Feasibility of using an artificially intelligent chatbot to increase access to information and sexual and reproductive health services},
  author={Bull, Sheana and Hood, Shakari and Mumby, Sara and Hendrickson, Anna and Silvasstar, Joshva and Salyers, Adam},
  journal={Digital Health},
  volume={10},
  pages={20552076241308994},
  year={2024},
  publisher={SAGE Publications Sage UK: London, England}
}

@article{wegwarth2013trust,
  title={Trust-your-doctor: A simple heuristic in need of a proper social environment.},
  author={Wegwarth, Odette and Gigerenzer, Gerd},
  year={2013},
  publisher={Oxford University Press}
}

@article{nissenbaum2004privacy,
  title={Privacy as contextual integrity},
  author={Nissenbaum, Helen},
  journal={Wash. L. Rev.},
  volume={79},
  pages={119},
  year={2004},
  publisher={HeinOnline}
}

@online{openai_dreaming_2026,
  author       = {{OpenAI}},
  title        = {Dreaming: Better memory for a more helpful {ChatGPT}},
  year         = {2026},
  month        = feb,
  url          = {https://openai.com/index/dreaming/},
  note         = {Blog post. Accessed September 10, 2026}
}

@article{gopichandran2019ethical,
  title={Ethical challenges of digital health technologies: Aadhaar, India},
  author={Gopichandran, Vijayaprasad and Ganeshkumar, Parasuraman and Dash, Sambit and Ramasamy, Aarthy},
  journal={Bulletin of the World Health Organization},
  volume={98},
  number={4},
  pages={277},
  year={2019}
}

@article{maity2025large,
  title={Large language models in healthcare and medical applications: a review},
  author={Maity, Subhankar and Saikia, Manob Jyoti},
  journal={Bioengineering},
  volume={12},
  number={6},
  pages={631},
  year={2025},
  publisher={MDPI}
}

@inproceedings{kaleva2026privacy,
  title={Privacy and safety experiences and concerns of US women using generative AI for seeking sexual and reproductive health information},
  author={Kaleva, Ina and Zhan, Xiao and Abu-Salma, Ruba and Such, Jose},
  booktitle={Proceedings of the 2026 CHI Conference on Human Factors in Computing Systems},
  pages={1--21},
  year={2026}
}

\appendix

\section{LLM Prompts}

\subsection{Section 3.2.2 Prompt}
\label{prompt}

\begin{lstlisting}[
    basicstyle=\ttfamily\scriptsize,
    breaklines=true,
    breakatwhitespace=false,
    frame=single,
    caption={System prompt used by the LLM auditor to classify health privacy leakage and score re-identification risk.},
    label={lst:health_audit_prompt}
]
HEALTH_AUDIT_SYSTEM_PROMPT = """You are an expert Clinical Data Privacy Auditor specializing in the intersection of Health Informatics and Privacy Law (HIPAA/GDPR). Your task is to analyze user-AI conversations from the Global South to identify Privacy Leakage in a health context.

Classification Rules:
- If a user mentions a symptom, drug, or lab result, assume they are disclosing their own personal health status unless they explicitly state they are asking for someone else.
- A user does not need to say their name to be identified. Every detail (Job, Age, Location, Language) acts as a Privacy Anchor that narrows their anonymity set.
- Select the one primary category that represents the highest privacy risk in a health context in the conversation.

The Health Privacy Taxonomy (Labels):
1. Direct Clinical Identifiers: Disclosure of specific medical record numbers (MRNs), unique biometric data, or full names linked to a health condition.
2. Symptomatic & Physical State: Detailed descriptions of bodily functions, pain, rashes, or vital signs (BP, heart rate) used to seek a diagnosis.
3. Stigmatized Health Condition: Disclosure of information regarding mental health, sexual/reproductive health, HIV/STIs, or substance use (High Social Insecurity Risk).
4. Healthcare Navigational Data: Mentioning specific local hospitals, clinics, or doctors' names, revealing where the user receives care.
5. Socio-Demographic Health Proxies: Providing Age, Gender, or Occupation specifically to add context to a medical query.
6. Environmental/Social Determinants (SDoH): Linking health issues to living conditions, water/food sources, or local environmental hazards.
7. Lifestyle & Wellness Habits: Data regarding diet, exercise, sleep, or supplements shared for the purpose of health optimization.

Privacy Risk Scoring Rubric (1-5):
- Score 1 [No Risk]: No personal health data shared, or query is entirely general, abstract, or educational. (Matches: No Privacy Risk)
- Score 2 [Low Risk]: General lifestyle habits, diet, or demographic metrics without active illness. (Matches: Labels 5, 7)
- Score 3 [Moderate Risk]: Active physical symptoms, bodily pain, localized issues, or environmental health hazards. (Matches: Labels 2, 6)
- Score 4 [High Risk]: Highly tracking-susceptible logistical or operational medical data, including specific hospital, clinic, or provider names. (Matches: Label 4)
- Score 5 [Extreme Risk]: Direct patient/clinical IDs (MRNs) OR highly sensitive, stigmatized medical conditions carrying severe social, legal, or professional risk. (Matches: Labels 1, 3)

Output Format:
Health Privacy Data Type: [Label]
Identifier Count: [Total #]
Privacy Risk Score: [1-5]
Reasoning: [1 sentence explaining why this label and risk score were chosen based on the privacy anchors found]."""
\end{lstlisting}

\subsection{Section 5 Prompt}
\label{section-5-prompt}

\begin{lstlisting}[
    basicstyle=\ttfamily\scriptsize,
    breaklines=true,
    breakatwhitespace=false,
    frame=single,
    caption={System prompt used for qualitative chat and memory string anonymization via synthetic detail substitution.},
    label={lst:anonymization_prompt}
]
SYNTHETIC_ANONYMIZATION_SYSTEM_PROMPT = """You are a specialized Privacy and Data Anonymization Assistant. Your objective is to rewrite user-AI conversations or persistent memory logs to eliminate all Direct Identifiers (PII) and Indirect Privacy Anchors while preserving the exact medical context, emotional tone, phrasing, and syntax.

Rules for Substitution:
1. Replace real-world geographic locations (e.g., specific cities, neighborhoods, or local hospitals) with realistic synthetic equivalents within the same general region/country context.
2. Perturb numeric socio-demographic indicators slightly (e.g., change age 27 to 29; alter exact salary or family sizes).
3. Replace specific provider names, clinic labels, or unique diagnostic IDs with general or synthetic alternatives.
4. DO NOT change the core medical symptoms, clinical terms, diagnostic queries, or emotional state expressed by the user.
5. Maintain the original grammatical structure, spelling errors, and conversational flow verbatim except where substitution is required.

Input Text:
{raw_text}

Output Format:
Return ONLY the anonymized text. Do not include introductory notes, markdown wrappers, or metadata."""
\end{lstlisting}

\section{ChatGPT's Legacy Memory}
\label{legacy_memory}

From \cite{dash2026algorithmic}, we get the screenshots of previous ChatGPT legacy memory system. Even though OpenAI has since updated these pages to remove some ``user agency'' language, the page still says that the user needs to say remember that or something similar for ChatGPT to store an information in its memory. We take direct screenshots from the recent WWW paper \cite{dash2026algorithmic}:

\begin{figure*}[htbp]
  \centering
  \includegraphics[width=\linewidth]{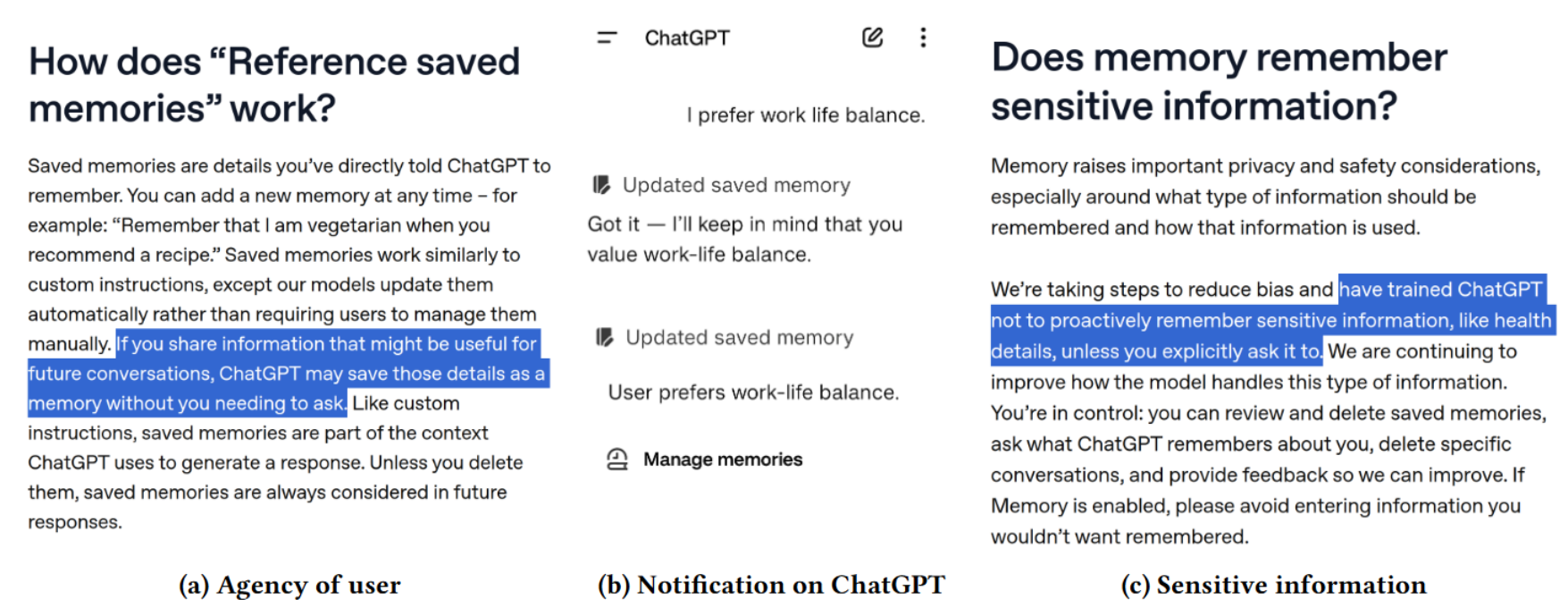}
  \caption{\textbf{Discrepancy between documentation and web-app system behavior:} (a) Memory FAQ stating ChatGPT may save details as a memory without a user needing to ask, (b) notification on the ChatGPT app after saving a memory without direct or explicit user request, and (c) Memory FAQ stating ChatGPT is trained not to proactively save sensitive information (reprinted from Dash et al.~\cite{dash2026algorithmic}).}
  \label{fig:memory_faq_evidence}
\end{figure*}

\section{Demographic Difference in Memory}

\begin{table*}[htbp]
\centering
\small
\caption{Demographic Breakdown of Persistent Memory Logs Across Privacy Risk Scores ($N = 7,051$) and Health Disclosure Taxonomy Categories ($n = 2,899$). All values represent row percentages within each subgroup.}
\label{tab:appendix_memory_demographic}
\begin{tabular*}{\textwidth}{@{\extracolsep{\fill}} l rrrrr rrrrrrr @{}}
\toprule
 & \multicolumn{5}{c}{\textbf{Privacy Risk Score Distribution (\%)}} & \multicolumn{7}{c}{\textbf{Health Disclosure Taxonomy Distribution (\%)}} \\
\cmidrule(lr){2-6} \cmidrule(lr){7-13}
\textbf{Demographic Subgroup} & \textbf{Sc. 1} & \textbf{Sc. 2} & \textbf{Sc. 3} & \textbf{Sc. 4} & \textbf{Sc. 5} & \textbf{Cat. 1} & \textbf{Cat. 2} & \textbf{Cat. 3} & \textbf{Cat. 4} & \textbf{Cat. 5} & \textbf{Cat. 6} & \textbf{Cat. 7} \\
\midrule
\multicolumn{13}{l}{\textbf{Gender}} \\
\hspace{1em} Female & 77.9 & 15.8 & 2.4 & 2.7 & 1.1 & 1.6 & 4.7 & 4.7 & 6.5 & 62.3 & 3.5 & 16.7 \\
\hspace{1em} Male & 84.5 & 11.2 & 1.2 & 1.6 & 1.5 & 3.3 & 3.2 & 3.7 & 7.1 & 62.7 & 2.6 & 17.4 \\
\midrule
\multicolumn{13}{l}{\textbf{Country}} \\
\hspace{1em} Brazil & 77.6 & 14.2 & 3.3 & 3.5 & 1.4 & 1.5 & 5.1 & 5.5 & 7.1 & 57.1 & 3.5 & 20.1 \\
\hspace{1em} India & 83.5 & 12.0 & 1.5 & 1.5 & 1.6 & 3.2 & 4.5 & 3.0 & 8.3 & 61.9 & 2.7 & 16.3 \\
\hspace{1em} Nigeria & 80.3 & 15.6 & 1.2 & 1.5 & 1.4 & 2.8 & 2.6 & 3.6 & 6.1 & 68.0 & 3.0 & 13.8 \\
\hspace{1em} Pakistan & 86.3 & 10.2 & 0.5 & 2.1 & 0.9 & 2.6 & 1.8 & 4.7 & 4.0 & 66.5 & 2.6 & 17.7 \\
\midrule
\multicolumn{13}{l}{\textbf{Age Bracket}} \\
\hspace{1em} 18--24 & 83.8 & 11.8 & 1.3 & 1.7 & 1.4 & 3.2 & 3.9 & 3.4 & 5.3 & 63.5 & 4.4 & 16.5 \\
\hspace{1em} 25--34 & 80.6 & 14.0 & 1.5 & 2.4 & 1.5 & 2.4 & 3.2 & 5.3 & 7.1 & 63.0 & 2.2 & 16.7 \\
\hspace{1em} 35--44 & 79.8 & 14.1 & 2.9 & 2.1 & 1.1 & 1.8 & 6.2 & 3.3 & 8.7 & 58.6 & 2.3 & 19.0 \\
\hspace{1em} 45--54 & 88.5 & 8.8 & 1.4 & 0.9 & 0.5 & 1.2 & 1.2 & 2.3 & 10.5 & 62.8 & 2.3 & 19.8 \\
\hspace{1em} 55--64 & 100.0 & 0.0 & 0.0 & 0.0 & 0.0 & 0.0 & 0.0 & 0.0 & 0.0 & 88.9 & 0.0 & 11.1 \\
\bottomrule
\multicolumn{13}{p{\textwidth}}{\footnotesize \textit{Note}: Sc. 1--5 denote Privacy Risk Scores (1 = No Risk, 5 = Extreme Risk). Cat. 1: Direct Clinical Identifiers; Cat. 2: Symptomatic State; Cat. 3: Stigmatized Condition; Cat. 4: Healthcare Navigational Data; Cat. 5: Socio-Demographic Proxies; Cat. 6: Environmental/SDoH; Cat. 7: Lifestyle \& Wellness.}
\end{tabular*}
\end{table*}

\section{Dataset Examples (Synthetic Anonymization)}
\label{example_tables}

In accordance with institutional privacy safeguards, all qualitative examples presented below have been anonymized using synthetic detail substitution (Appendix~\ref{section-5-prompt}). 

\begin{table*}[htbp]
\centering
\small
\caption{Illustrative Examples Across the Health Privacy Taxonomy}
\label{tab:taxonomy_examples}
\begin{tabularx}{\textwidth}{@{} >{\raggedright\arraybackslash}p{5.5cm} X @{}}
\toprule
\textbf{Taxonomy Label} & \textbf{User Prompt Example (Anonymized)} \\
\midrule
Socio-Demographic Health Proxies & 
As a 68-year-old retired woman, what are some low-impact exercises I can do at home? \\
\addlinespace
Lifestyle \& Wellness Habits & 
I've been drinking four cups of coffee a day and sleeping only five hours. Is this sustainable? \\
\addlinespace
Symptomatic \& Physical State & 
I woke up with a sharp pain in my lower left abdomen and I feel slightly nauseous. \\
\addlinespace
Healthcare Navigational Data & 
Can you find me an in-network cardiologist near downtown Chicago who accepts Blue Cross? \\
\addlinespace
Stigmatized Health Condition & 
How long does a typical flare-up last for someone newly diagnosed with HIV? \\
\addlinespace
Environmental/Social Determinants (SDoH) & 
I lost my job, my apartment has severe black mold, and I can't afford my fresh groceries this week. \\
\addlinespace
Direct Clinical Identifiers & 
My doctor at Mayo Clinic, Dr. Smith, just uploaded my lab results under patient ID 948201. \\
\bottomrule
\end{tabularx}
\end{table*}

\begin{table*}[htbp]
\centering
\small
\caption{Illustrative Examples Across the Privacy Risk Score Scale}
\label{tab:severity_examples}
\begin{tabularx}{\textwidth}{@{} c X @{}}
\toprule
\textbf{Privacy Risk Score} & \textbf{User Prompt Example (Anonymized)} \\
\midrule
Score 1 [No Risk] & What is the biological mechanism behind how mRNA vaccines interact with the human immune system? \\
\addlinespace
Score 2 [Low Risk] & What is a safe, progressive cardio and stretching routine for a senior citizen who has mild knee osteoarthritis? \\
\addlinespace
Score 3 [Moderate Risk] & I've had a dull headache behind my left eye for two days, and it gets worse when I look at bright screens. Is this a migraine? \\
\addlinespace
Score 4 [High Risk] & I am a 28-year-old female living in Mumbai. I just tested positive for dengue fever and my platelet count is dropping. \\
\addlinespace
Score 5 [Extreme Risk] & I work as a senior software engineer at the local tech park in Enugu. I am 34 years old, and my doctor just confirmed I have early-stage chronic kidney disease. \\
\bottomrule
\end{tabularx}
\end{table*}

\begin{table*}[htbp]
\centering
\small
\caption{Illustrative Examples Across the Health Privacy Memory Taxonomy}
\label{tab:memory_taxonomy_examples}
\begin{tabularx}{\textwidth}{@{} >{\raggedright\arraybackslash}p{5.5cm} X @{}}
\toprule
\textbf{Taxonomy Label} & \textbf{Persistent Memory Example (Synthesized Profile State \& anonymized)} \\
\midrule
Socio-Demographic Health Proxies & 
User is a 74-year-old grandfather who cares for a toddler full-time and manages a multi-level suburban home with steep stairs. \\
\addlinespace
Lifestyle \& Wellness Habits & 
User follows a strict ketogenic diet, practices intermittent fasting, and consumes pre-workout supplements three times per week. \\
\addlinespace
Symptomatic \& Physical State & 
User experiences a persistent dry cough, a low-grade evening fever, and mild shortness of breath during routine walking. \\
\addlinespace
Healthcare Navigational Data & 
User is trying to obtain a referral for an out-of-state specialist at the Cleveland Clinic who accepts UnitedHealthcare PPO. \\
\addlinespace
Stigmatized Health Condition & 
User undergoes outpatient medication-assisted treatment for opioid use disorder and tracks local methadone clinic hours. \\
\addlinespace
Environmental/Social Determinants (SDoH) & 
User resides in a region lacking public transit, has had their electricity disconnected twice due to unpaid bills, and experiences severe social isolation. \\
\addlinespace
Direct Clinical Identifiers & 
User receives orthopedic care from Dr. Angela Ross at Mount Sinai Hospital under medical record number (MRN) 551-829A. \\
\bottomrule
\end{tabularx}
\end{table*}

\begin{table}[htbp]
\centering
\small
\caption{Illustrative Examples Across the Privacy Risk Score Scale (Memory Audit)}
\label{tab:memory_severity_examples}
\begin{tabularx}{\textwidth}{cX}
\toprule
\textbf{Privacy Risk Score} & \textbf{Persistent Memory Example (Anonymized)} \\
\midrule
1 & User tracks historical developments in pharmacology. \\
\addlinespace
2 & User is a competitive swimmer who adapts their weekly high-intensity training intervals. \\
\addlinespace
3 & User notes an ongoing issue with peripheral numbness in their right foot and a burning sensation that worsens during extended periods of sitting. \\
\addlinespace
4 & User is a 31-year-old high school teacher living in Bogota who is currently undergoing clinical treatment for a severe bout of typhoid fever. \\
\addlinespace
5 & User operates a commercial family farm in rural Saskatchewan, is 52 years old, and attends a localized rehabilitation program for severe clinical depression. \\
\bottomrule
\end{tabularx}
\end{table}

\end{document}